\documentclass[useAMS,usenatbib]{mn2e}

\usepackage{times}
\usepackage{graphicx}
\usepackage{rotating}

\def\pcm3{{\rm\thinspace cm^{-3}}}

\def\n_h{{\rm n_{H}}}

\def\NH1{{$N_{\rm HI}~$}}

\def\ga{{\rm\thinspace gauss}}

\def\mic{{\rm\thinspace $\mu$m}}

\def\approxlt{\mathrel{\hbox{\rlap{\lower .5ex \hbox {$\sim$}}
        \raise .15 ex \hbox{$<$}}}}
\def\approxgt{\mathrel{\hbox{\rlap{\lower .5ex \hbox {$\sim$}}
        \raise .15 ex \hbox{$>$}}}}

\def\la{\mathrel{\hbox{\rlap{\hbox{\lower4pt\hbox{$\sim$}}}\hbox{$<$}}}}
\def\ga{\mathrel{\hbox{\rlap{\hbox{\lower4pt\hbox{$\sim$}}}\hbox{$>$}}}}
\newbox\grsign \setbox\grsign=\hbox{$>$} \newdimen\grdimen
\grdimen=\ht\grsign
\newbox\simlessbox \newbox\simgreatbox \newbox\simpropbox
\setbox\simgreatbox=\hbox{\raise.5ex\hbox{$>$}\llap
     {\lower.5ex\hbox{$\sim$}}}\ht1=\grdimen\dp1=0pt
\setbox\simlessbox=\hbox{\raise.5ex\hbox{$<$}\llap
     {\lower.5ex\hbox{$\sim$}}}\ht2=\grdimen\dp2=0pt
\setbox\simpropbox=\hbox{\raise.5ex\hbox{$\propto$}\llap
     {\lower.5ex\hbox{$\sim$}}}\ht2=\grdimen\dp2=0pt

\title[Methane and Spitzer imaging of Pleiades dwarfs]{Methane band and Spitzer mid-IR imaging of L and T dwarf candidates in the Pleiades}
\author[S. L. Casewell et al.]{S. L. Casewell$^{1}$\thanks{E-mail:
slc25@star.le.ac.uk},  R. F. Jameson$^{1}$, M. R. Burleigh $^{1}$,  P. D. Dobbie$^{2}$, M. Roy$^{1}$,
\newauthor S. T. Hodgkin$^{3}$ and E. Moraux $^{4}$
\\ 
$^{1}$Department of Physics and Astronomy, University of Leicester, University Road, Leicester LE1 7RH, UK\\
$^{2}$Australian Astronomical Observatory, PO Box 296, Epping NSW 1710 Australia\\
$^{3}$CASU, Institute of Astronomy,University of Cambridge, Maddingley Road, Cambridge, CB3 0HA, UK \\
$^{4}$Laboratoire d'Astrophysique, Observatoire de Grenoble, Universit\'e Joseph Fourier, BP 53, 38041 Grenoble Cedex 9, France\\
}
\begin{document}

\date{\today}

\pagerange{\pageref{firstpage}--\pageref{lastpage}} \pubyear{2007}

\maketitle

\label{firstpage}

\begin{abstract}

We present $Spitzer$ observations at [3.6] and [4.5] microns together with the methane short (1.58\,\mic)-methane long (1.69\,\mic) colour for 3 cool dwarfs
in the Pleiades, PLZJ23, PLZJ93 and PLZJ100.
We determine the effective temperatures of PLZJ23 and PLZJ93 to be $\approx$1200 and $\approx$1100\,K. From the broadband photometry we place an upper limit of 1100 K on the effective temperature of PLZJ100 but lack the 
data required to determine the value more precisely. These temperatures are in the T dwarf regime yet the methane colours indicate no methane is present.We attribute this to youth/low gravity in line with theoretical expectations. However, we find even less methane is present than predicted by the models. 

PLZJ23 and PLZJ93 are also very bright in the [3.6] micron waveband (PLZJ100 is not measured) compared to field brown dwarfs which can also be explained by this lack of methane.

 The definition of the T spectral class is the appearance of methane absorption, so strictly, via this definition, PLZJ93 and PLZJ100 cannot be described as T dwarfs. The colours of these two objects are, however, not compatible with those of L dwarfs. Thus we have a classification problem and cannot 
assign these objects a (photometric) spectral type.

\end{abstract}


\section{Introduction}

Since the discovery of the first T dwarf, Gl229B \citep{nakajima95}, more than 150 have been catalogued \footnote[1]{http://spider.ipac.caltech.edu/staff/davy/ARCHIVE/index.shtml}. These are almost exclusively field dwarfs, and have been used to define the T spectral type and to produce empirical sequences that describe how the photometric colours change with effective temperature (e.g. \citealt{leggett02,leggett10}). However, as brown dwarfs are degenerate objects, their radii decrease as they cool \citep{burrows01}, making not only this parameter, but also their gravity permanently linked to their evolution. There has been much work on searching for both T and, the earlier spectral type, L dwarfs in open star clusters as cluster members have a known age. This age constraint means that evolutionary models (e.g. DUSTY, \citealt{chabrier00}) can be used to estimate their masses and, as the majority of cluster members remain gravitationally bound, the mass function of the cluster can be used as a proxy for the initial mass function (IMF) of the cluster, and hence the Galaxy. There is much discussion as to the shape of the IMF and whether it is consistent across all open star clusters (e.g. \citealt*{moraux05}; \citealt{lodieu07}). Inconsistencies would suggest a variety of initial conditions to star formation processes. The IMF can also be used to put constraints on the lowest masses that can be created via star formation processes. The majority of the work on the IMF to date has been based on L dwarfs alone. To better understand the IMF we must begin to discover more T dwarfs with a known age, thus extending the IMF into the T dwarf regime.This is because only with a sample that encompasses a wide range in mass and age can we begin to fully comprehend the IMF. 

The majority of discovered T dwarfs however,  are field objects, and hence old ($>$5 Gyr). This means that they have no accurate age estimation to enable them to be used in defining the IMF, and  empirical relations derived from colours and spectra provide little information on how gravity affects their atmospheres.
To date there are few examples of confirmed low gravity T dwarfs (e.g. HN Peg B \citealt{luhman07}; SOri70,  \citealt{zapatero02}) and one object on the L-T transition HD203030B \citep*{metchev06}. The L-T transition is the region where the dusty L dwarf atmospheres begin to clear, and the object's atmosphere becomes dominated by methane absorption, the defining characteristic of T dwarfs.  The processes involved are not yet well understood, and the majority of evolutionary models do not cover this region, instead having a ``dusty'' set of models (e.g. DUSTY, \citealt{chabrier00}; C100, \citealt{burrows06}) and a clear atmosphere set of models (e.g. COND, \citealt{baraffe03}; Clr, \citealt{burrows06}) for the later T dwarfs. \citet*{saumon08} attempted to reproduce this region using a hybrid model for the colour evolution and cooling of the brown dwarfs. They had some success in reproducing the photometric characteristics of the field population, but made less progress with the younger  members of the Pleiades cluster.
These currently known low-gravity objects have been identified as such thanks to fiducial age constraints demonstrating the youth
of the systems. Such objects represent benchmarks as with a known age (either from their being a cluster member or in a binary with a star/ white dwarf of a known age), they can be used to critically examine the predictions made by evolutionary models.

Brown dwarfs are classified as either L or T dwarfs based on the strengths of lines within their spectra, and the general spectral shape in comparison to standards (see \citealt{kirkpatrick05} for a review), although estimates based on broadband photometry are often made. For T dwarfs this classification is usually based on the near-infrared spectra \citep{burgasser02, geballe02}, while the hotter L dwarfs are classified using optical spectra \citep{kirkpatrick99}. Obviously, to achieve any sort of classification, ``standard'' objects must be selected, and so objects with unusual colours such as low gravity objects were often excluded as outliers. 
\citet{kirkpatrick05} notes that, for example,  any L0 dwarf will fall between a higher mass ($\approx$85 M$_{\rm Jup}$), old ($>$ a few Gyr), stellar object to a very young ($<$ 20 Myr) low mass ($<$20 M$_{\rm Jup}$) brown dwarf. Understandably, this is a large range in mass and age, and it is therefore perhaps naive to assume every L dwarf of a defined spectral type to  have identical photometry and indeed spectra, particularly while so many absorption lines are sensitive to gravity (e.g. K \textsc{i}, Na \textsc{i}, Rb \textsc{i}, Cs \textsc{i}). It was suggested in \citet{kirkpatrick05} that brown dwarf classification have a suffix added where $\alpha$, $\beta$, $\gamma$ and $\delta$ could be used to denote decreasing gravity. \citet{cruz09} have recently used this notation while creating a low gravity sequence for early L dwarfs. This follows the work of 
\citet{kirkpatrick08} who studied 20 low gravity dwarfs including 8 M dwarfs and 11 L dwarfs. \citet{cruz09} identified 23 low gravity L dwarfs from L0-L5, and beginning to define  a low gravity sequence based on that presented in \citet{kirkpatrick99} using gravity sensitive features in the spectra.
They use $\beta$ to describe an object of the age of the Pleiades, and show that typically these objects are red for their spectral class in $J-K_{S}$. It will, however, take time before a similar feat can be accomplished based on late L dwarfs and T dwarfs, as so few are currently known. 

As discussed earlier, open star clusters have long been the favourite hunting ground for these benchmark objects as they are coeval. However, as we search further into the T dwarf regime to cooler temperatures, it is becoming evident that the majority of surveys do not probe deep enough to discover T dwarfs. To date, there are 2 T dwarfs known in the Hyades \citep{bouvier08}, although at an age of 625 Myr this means that they have a gravity similar to that of field T dwarfs. SOri70 \citep{zapatero02} was discovered in $\sigma$ Orionis, and it has been recently suggested that it may be surrounded by a dust disk accounting for its redness in the [3.6] and [4.5] micron bands \citep{scholz08}. One object was reported by \citep{burgess09} in IC348 with an estimated spectral type of T6. However at the young age of this population (3 Myr), the reliability of evolutionary models is questionable
due to the rather ad-hoc nature of the starting point of these
calculations \citep{baraffe02}. 

 There have been many studies of the Pleiades to search for young brown dwarfs, although until our previous work \citep{casewell07} only L dwarfs were discovered (e.g. \citealt{bouvier98, pinfield00, moraux01, dobbie02, bihain06, stauffer07, lodieu07}).
In \citet{casewell07} we presented the results of a 2.5 square degree survey of the Pleiades in near-IR and optical wavebands. After selecting candidate cluster members
using the $I-Z$, $I$ and $Z-J$, $Z$ colour magnitude diagrams, we measured proper motions using the 5 year baseline between the near-IR and optical datasets. After our analysis, 6 candidates were consistent with being late L or early T dwarfs. These objects are far too faint to obtain a spectrum with which to confirm their nature, and so we present here additional photometry in the form of $Spitzer$ IRAC [3.6] and [4.5] micron bands, methane imaging and near-IR $K$ band imaging of the faintest candidates. This additional information will allow us to place constraints on atmospheric models appropriate to a young cluster and hence a low gravity environment.

\section{Observations and  Data Reduction }

\subsection{The methane imaging and its reduction}

\begin{figure}
\begin{center}
\scalebox{0.35}{\includegraphics[angle=270]{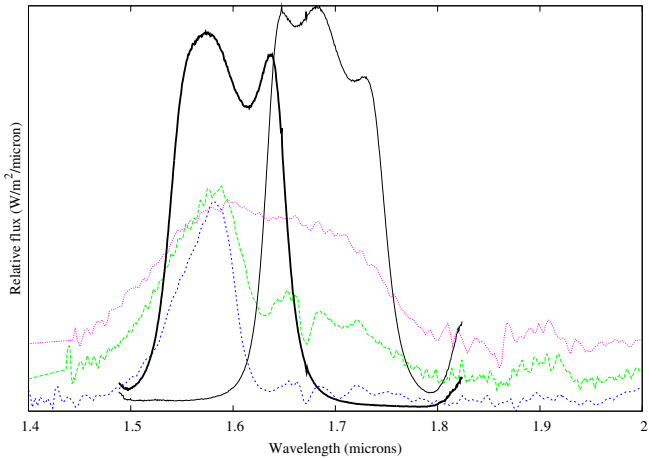}}
\caption{\label{meth_flt}The $H$ passband ($\approx$1.45-1.85\,\mic) with the $CH_{4}$ short filter (thick black line), $CH_{4}$ long filter (thin black line) and three field T dwarfs,  SDSSJ015141.69+124429.6 (T0.5, magenta, dotted line),  J05591914-1404488 (T4.5, green, long dashed line) and 2MASSJ0415195-093506 (T8, blue, dashed line). The effect of methane absorption is clearly seen in the $H$ band of the later T dwarfs.
}
\end{center}
\end{figure}

Methane imaging is a very useful tool for determining whether a brown dwarf is a bona fide T dwarf when a spectrum is not available as shown by \citet{tinney05}. 
As T dwarfs get cooler, methane absorption becomes more dominant in the $H$ band as seen in Figure \ref{meth_flt}. The magenta curve (long dashed line) is 
a T0.5 dwarf, SDSSJ015141.69+124429.6 \citep{geballe02} which shows very little methane absorption. The flux in the $H$ band forms a broad hump. However, as we progress to the T4.5 dwarf, 2MASSJ05591914-1404488 (green, short dashed line; \citealt{burgasser00}) this region, and in particular the red side of the flux,  appears to have been suppressed by the presence of methane in the atmosphere. This is further shown in the T8 dwarf, 2MASSJ0415195-093506 (blue, dotted line; \citealt{burgasser02})  which shows sharply peaked flux in this region.
We can further characterise the degree of methane absorption by using the methane filters,  $CH_{4}$ ``long'' (grey line in Figure 
\ref{meth_flt}) and ``short'' (black line in Figure \ref{meth_flt}). We can  calculate a methane colour index $CH_{4s}$-$_{4l}$, which 
will be negative for objects with a late T spectral type as all the flux is in the short filter and not the long (where the majority of the methane absorption is occurring).

We imaged each of the 5  T dwarf candidates from \citet{casewell07} (PLZJ93, PLZJ721, PLZJ235, PLZJ112 and PLZJ100)  using NIRI on Gemini North in both the methane short and methane long filters on the nights of 02/09/2007 and 04/09/2007. We requested that the seeing be better than 1.2'' and that the cloud cover was at the 70th percentile or better (corresponding to at worst, thin cirrus). It was noted in the observing log there was some passing cloud on the observation of PLZJ100, but this does not appear to have affected the images.  The remaining candidate PLZJ23, has an estimated photometric spectral type of L8, and therefore was not observed as part of the methane imaging programme.

NIRI consists of a 1024$\times$1024 pixel ALADDIN InSb array and when combined with the f/6 camera, provides a plate scale of 0.117 arcseconds per pixel and a field of view of 120$\times$120 arcseconds. 
The data were obtained using a 9 point dither pattern and total exposure times of 30 mins for the $CH_{4s}$ filter and 1hr for the $CH_{4l}$ filter. We also observed a known T5 dwarf 2MASSJ04070885+1514565 \citep{burgasser04} for 1 min in both the $CH_{4s}$ and $CH_{4l}$ filters to act as a standard object with which to check our results.

\begin{figure}
\begin{center}
\scalebox{0.3}{\includegraphics[angle=270]{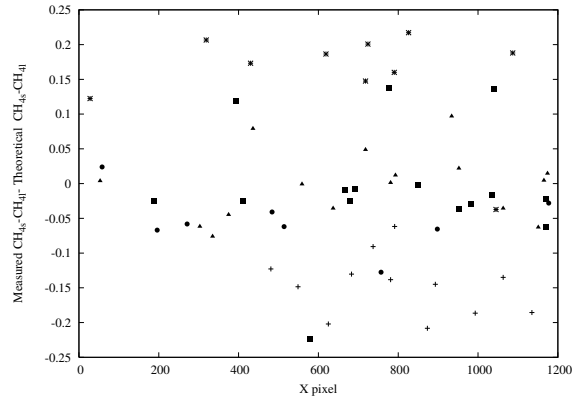}}
\caption{\label{xpix}Difference in zeropoint for each reference object used per target (PLZJ93 with crosses, PLZJ100 with circles, PLZJ112 with asterisks, PLZJ235 with triangles and  PLZJ721 with boxes) with X pixel across the NIRI chip.
}
\end{center}
\end{figure}

\begin{figure}
\begin{center}
\scalebox{0.3}{\includegraphics[angle=270]{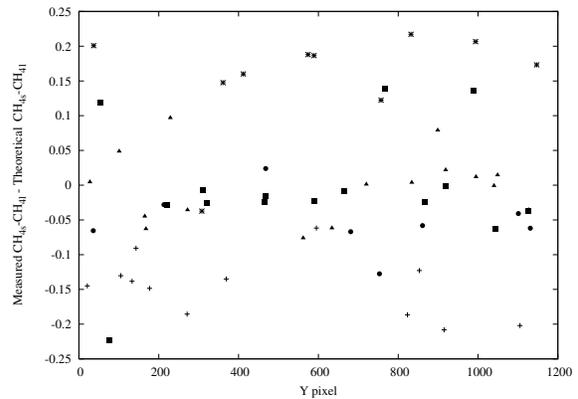}}
\caption{\label{ypix}Difference in zeropoint for each reference object used per target (PLZJ93 with crosses, PLZJ100 with circles, PLZJ112 with asterisks, PLZJ235 with triangles and  PLZJ721 with boxes) with Y pixel across the NIRI chip. 
}
\end{center}
\end{figure}

The images were reduced using  \textsc{iraf}, the \textsc{gemini} package v1.9 and the NIRI specific tasks.
Firstly the images were prepared using the \textsc{nprepare} task which adds keywords to the image headers in preparation for data reduction. A sky frame was then created from the images using \textsc{nisky}, which median combined each science image after masking out any objects. This sky frame was used instead of dark frames, as having been taken concurrently with the science images, it gives a better measurement of the dark current. The \textsc{nireduce} task was then used to subtract the sky frame from the science images.
Flat field images are taken for NIRI in both a ``lamp on `` and a ``lamp off'' mode. These are essentially high and low dome flats and are used to separate the instrument thermal signature from the instrument response. The flat field was created by subtracting the median combined ``lamp off'' frames from the single ``lamp on'' frames and median combining. This flat field was then normalised by dividing by the mean pixel value with \textsc{niflat}. The \textsc{xdimsum} package and the \textsc{xmosaic} task were then used to combine the individual flat fielded science frames. 
This processing resulted in two images for each object, a $CH_{4l}$ and a $CH_{4s}$ image.
The images were then astrometrically calibrated using the \textsc{starlink} packages \textsc{autoastrom} and \textsc{gaia}, and the object extraction routine \textsc{SExtractor} was run. 
All sources in the images with good photometry were then selected, and $J$ and $K$ band photometry obtained for them from the UKIRT Infrared Sky Survey's Galactic Cluster Survey (UKIDSS GCS; \citealt{lawrence07}) Data Release 6.
Stellar sources were then selected using the flags from UKIDSS and their $J-K$ colours were then used with Figure 4 of \citet{tinney05} to estimate the spectral type for these reference objects (typically between 12 and 20 objects were used). As NIRI and IRIS2 have the same methane filter set Equation 2 from \citet{tinney05} was then used to calculate the theoretical $CH_{4s}$ -$CH_{4l}$ colour. The measured $CH_{4s}$ -$CH_{4l}$ were then plotted against the theoretical colour and used to determine the zeropoint for the data. This zeropoint was then applied to the science targets. The zeropoints for each of the reference objects used are shown in figures \ref{xpix} and \ref{ypix}, and the scatter in the standards is typically about 10 per cent.

We checked the reliability of our method by using 2MASSJ04070885+1514565. We had fewer standards, as the UKIDSS LAS has not covered this region yet, and so we used 2MASS with the transforms presented by \citep{carpenter01} to convert between the 2MASS and MKO filter set. The resultant $CH_{4s}$ -$CH_{4l}$ colour was -0.569$\pm$0.0293 which is consistent with colours for a T5 dwarf.

\begin{figure}
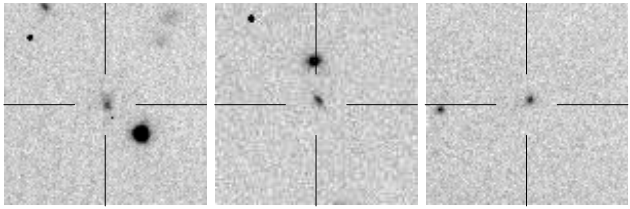

\begin{center}
\scalebox{0.6}{\includegraphics[angle=270]{plzj112.ps}}
\scalebox{0.6}{\includegraphics[angle=270]{plzj235.ps}}
\scalebox{0.6}{\includegraphics[angle=270]{plzj721.ps}}

\caption{\label{galaxy}40'' square Gemini NIRI images of PLZJ112 (left), PLZJ235 (centre) and PLZJ721 (right). The target is marked by the cross. All three of these targets appear to be background galaxies, and hence not members of the Pleiades.
}
\end{center}
\end{figure}

The methane images had much higher S/N than the original detection images, and in some cases, this means we were able to identify our targets as non-stellar objects. 
On close examination of the images, PLZJ235 appears to be a galaxy with a measured ellipticity of 1.53 (Figure \ref{galaxy}). PLZJ112 appears to be a galaxy, but is much less elongated than PLZJ235, and only has an ellipticity of 0.28. PLZJ721 does not appear to be a galaxy, but it has a measured  ellipticity of 2.02. However it has a methane colour index of 0.25$\pm$0.09 which is consistent with both PLZJ235 and PLZJ112 and  is also slightly offset from its position as measured by our original WFCAM images. This indicates that in our original analysis \citep{casewell07} it may only have been a marginal detection of part of an elliptical object. This would also account for the previous motion derived.

\begin{figure}
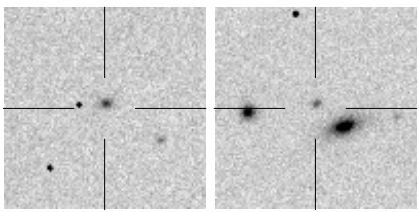

\begin{center}
\scalebox{0.6}{\includegraphics[angle=270]{plzj93.ps}}
\scalebox{0.6}{\includegraphics[angle=270]{plzj100.ps}}
\caption{\label{star}40'' square Gemini NIRI images of PLZJ93 (left), PLZJ100(right). The target is marked by the cross. These two objects appear to be stellar, and hence members of the Pleiades.
}
\end{center}
\end{figure}

PLZJ93 and PLZJ100 however, appear to be stellar with ellipticities of 0.30 and 0.23 respectively (Figure \ref{star}). Using the NIRI images we have also attempted to measure proper motions for these objects, however the small field of view means that the errors are too large to be an improvement on the measurements presented in \cite{casewell07}. The proper motions, are however consistent with the earlier results.
The final magnitudes are shown in Table \ref{meth_table}. 

\begin{table*}
\caption{\label{meth_table}Observed magnitudes and colours  for the candidate Pleiads.
PLZJ721, PLZJ235 and PLZJ112 have been determined to be background galaxies, and hence non-members of the Pleiades.}
\begin{center}
\begin{tabular}{l c c c c c c c c}
\hline
Name & RA & Dec &$I$&$Z$&$J$&$H$&$K$&$CH_{4s}-CH_{4l}$\\
&\multicolumn{2}{c}{J2000.0}&&&&&&\\
\hline
PLZJ23&03 51 53.38&+24 38 12.11&23.541$\pm$0.140&22.187$\pm$0.112&19.960$\pm$0.100&19.362$\pm$0.100&18.510$\pm$0.030&-\\
PLZJ93&03 55 13.00 &+24 36 15.8&24.488 $\pm$   0.370  &     22.592 $\pm$ 0.164   &      19.968 $\pm$   0.080    & 	 19.955$\pm$0.100&19.420 $\pm$0.100&0.14$\pm$0.08\\	
PLZJ100&03 47 19.19&+25 20 53.3&-&23.563$\pm$0.373&20.254$\pm$0.114&-&19.899$\pm$0.06&0.115$\pm$0.096\\
PLZJ721&03 55 07.14&+24 57 22.34&-&22.195$\pm$0.092&20.248$\pm$0.116&20.417$\pm$0.123&-&0.259$\pm$0.0982\\
PLZJ235&03 52 32.57&+24 44 36.64&-&22.339$\pm$0.115&20.039$\pm$0.112&20.245$\pm$0.127&-&0.215$\pm$0.097\\
PLZJ112&03 53 19.37&+24 53 31.85&-&22.532$\pm$0.116&20.281$\pm$0.143&-&-&0.295$\pm$0.0930\\

\hline
\end{tabular}
\end{center}
\end{table*}

\subsection{The Spitzer imaging and its reduction}
We have also obtained $Spitzer$  IRAC \citep{fazio04} photometry  from Cycle 4 (Programme ID:40482,  PI: Jameson) in the [3.6] and [4.5] micron bands using IRAC mapping mode to observe all of our remaining targets (PLZJ23, PLZJ93 and PLZJ100) on 17/10/2007 and 21/09/2008.
PLZJ100 however, is too close to a much brighter object for us to measure its photometry and so the data presented here is for PLZJ23 and PLZJ93 only.
The data were observed using 100s integrations and the standard medium 18 point dither pattern and the full array giving a total exposure time of 1800s per pointing.
We used the basic calibrated data (BCD) images and the standard reduction method with the recommended \textsc{mopex} pipeline (v18.3.3) to perform overlap corrections, and create final and array correction mosaics of the images.

We performed aperture photometry on our targets using the \textsc{apex} software and an aperture of 3 pixels with a background aperture of 12-20 pixels. We then used the array correction mosaic to correct for the array location dependence and the relevant zeropoint. Channel 1 ([3.6] micron) also had a pixel phase correction applied. The IRAC zero magnitude flux densities as found on the $Spitzer$ website were then used to convert the flux into magnitudes on the Vega magnitude scale.
The photometric errors were estimated using the Poisson noise given by the  \textsc{apex} software which was then combined with the errors on the zeropoints. The 3 per cent absolute calibration was added in quadrature to these photometric errors. We confirmed this error estimate was accurate by creating 3 stacks of 6 individual images in each channel, to corroborate the measured range of magnitudes was within the previously calculated error range.

\begin{table}
\caption{\label{sp_table}$Spitzer$ IRAC, [3.6] and [4.5] micron magnitudes for the candidate Pleiades L and T dwarfs PLZJ23 and PLZJ93.}
\begin{center}
\begin{tabular}{l c c }
\hline
Name &[3.6]&[4.5]\\
\hline
PLZJ93&17.266$\pm$0.045&16.929$\pm$0.057\\	
PLZJ23&16.655$\pm$0.040&16.194$\pm$0.044\\
\hline
\end{tabular}
\end{center}
\end{table}

\subsection{Additional $H$ and $K$ band photometry}
\begin{figure}
\begin{center}
\scalebox{0.35}{\includegraphics[angle=270]{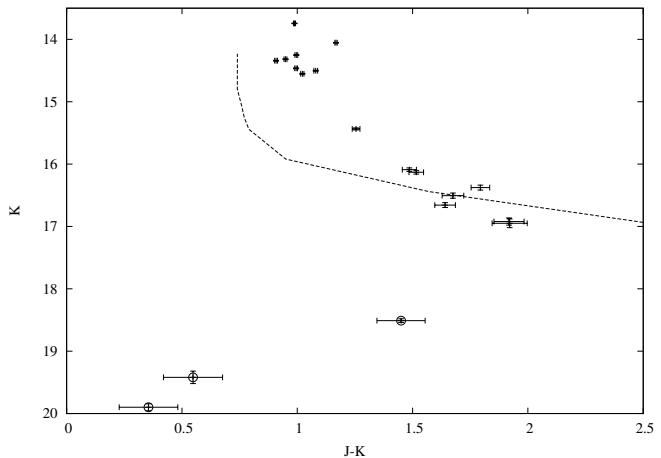}}
\caption{\label{jk}$J-K$, $K$ colour-magnitude diagram for all the candidate Pleiads with $K$ band measurements. The DUSTY model for 120 Myr is the dashed line. PLZJ23, PLZJ93 and PLZJ100 are the three faintest, and bluest objects and are marked by open circles.
}
\end{center}
\end{figure}

As our original data was in the $I$, $Z$, $J$ and in limited cases also the $H$ band, we  obtained time on UKIRT with the UKIRT Fast Track Imager (UFTI) in semester 2007B to observe our candidates in the $H$ and $K$ bands. Unfortunately bad weather meant that only $K$ band data were obtained for PLZJ100 on 26/01/2008. These data were reduced using the ORAC-DR pipeline provided as part of the  \textsc{starlink} packages, and the mosaics combined using the Image Reduction and Analysis Facility v2.12.2a (\textsc{iraf}; \citealt{tody86}) \textsc{xdimsum} package and the \textsc{xmosaic} task as described below. The  \textsc{SExtractor} was used to extract objects with stellar profiles from the images, and these objects were then cross correlated with the UKIDSS GCS catalogue to provide a photometric zeropoint and  calibrate the image.
The target's apparent magnitude is given in Table \ref{meth_table}, and the  $J-K$, $K$  colour-magnitude diagram for the cluster is shown in Figure \ref{jk}.

\section{Results}

Our new results leave three Pleiades candidates, PLZJ23, PLZJ93 and PLZJ100. Their broadband colours (Table \ref{meth_table}) indicate spectral types of late L for PLZJ23 and early to mid T for PLZJ93 and PLZJ100 and their position in the colour-magnitude diagrams and their proper motions are indicative of their being cluster members \citep{casewell07}. PLZJ93 and 100 also appear to have methane indices indicative of stellar objects and have low ellipticities on the images. Their methane colours(Table \ref{meth_table}) are not, however,  consistent with the spectral types that are suggested by the broadband colours and are more indicative of late L or early T types.  Additionally, this lack of methane absorption is not consistent with their being field dwarfs.

\begin{figure}
\begin{center}
\scalebox{0.35}{\includegraphics[angle=270]{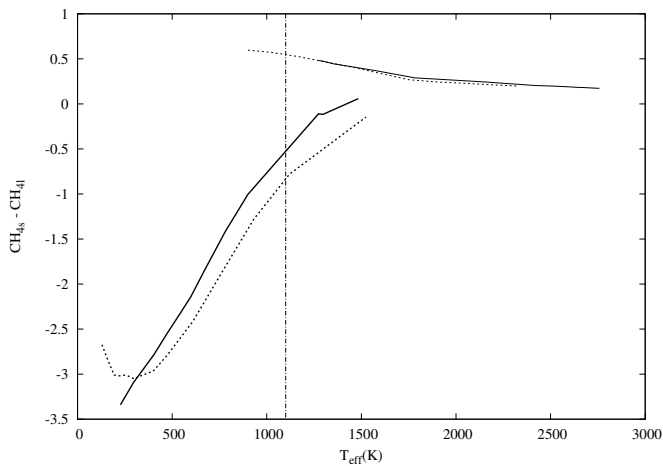}}
\caption{\label{models}$CH_{4s}$-$CH_{4l}$ colour vs T$_{\rm eff}$ for the DUSTY models at 5\,Gyr (thin dashed line) and 120 Myr (thin solid line) and the COND models at 5\,Gyr (thick dashed line) and 120\,Myr (thick solid line). The vertical dot-dashed line is at 1100\,K, the effective temperature of PLZJ93.}

\end{center}
\end{figure}

Figure \ref{models} shows the DUSTY \citep{chabrier00} and COND \citep{baraffe03} models that were generated in the NIRI methane filter set using the Phoenix web simulator (http://phoenix.ens-lyon.fr/simulator/index.faces). Both sets of models have been calculated for 5\,Gyr and 120\,Myr. It can be seen that we would expect less methane absorption at a given temperature for the lower gravity, younger Pleiades objects using the COND models (which are appropriate at this temperature) than for field objects at 5\,Gyr.

This result is not wholly surprising. \citet*{saumon08} created isochrones at the age of the Pleiades, but discovered that in order to make their model isochrones a better fit with the members of the Pleiades identified in \citet{casewell07}, they need to lower the effective temperature of the L-T transition which is set at 1200-1400\,K.  This suggests that for younger ages and hence lower gravities, the L-T transition may happen at a lower effective temperature. This is consistent with the analysis of HN Peg B presented by \citet{leggett08}.
 In IC348, \citet{burgess09} find that for younger objects, a cooler temperature is expected for the same methane colour. This is consistent with the assumption that the L-T transition, and hence the onset of methane absorption, occurs at a lower temperature for younger (lower gravity) objects.

The recent spectroscopic study on the Pleiades L dwarfs by \citet{bihain10}, also finds evidence of low gravity features in some early L dwarfs. These spectra show the characteristic triangular shaped bump in the in the $H$ band as seen in young L dwarfs in Trapezium \citep{lucas00}. This shape could be due to a reduction in the H$_{2}$ collision -induced absorption and water absorption in a low gravity and dustier environment than is commonly seen in field L dwarfs (\citealt*{borysow97}; \citealt{kirkpatrick06,mohanty07}). This dustier environment is consistent with the L-T transition happening at a lower temperature in the Pleiades. This youth and dusty atmosphere is also given as a reason for the redder $J-K$ colours seen in the Pleiades L dwarfs compared to field dwarfs as the H$_{2}$ collision -induced absorption opacity also affects the $K$ band and depends on the square of the gas density \citep{knapp04}, which is lower at lower gravity, making the atmosphere more transparent and hence the flux brighter.

A similar effect was seen by \citet{lodieu09} in Upper Scorpius where numerous candidates with spectral types between M and mid-L were selected from the UKIDSS GCS using photometry, but spectra revealed them to all have much earlier spectral types than expected. This lead the authors to deduce that the $J-K$ colour for young L dwarfs is much redder than older L dwarfs of the same spectral type.

To compare the PLZJ23, PLZJ93 and PLZJ100 with models, we have derived effective temperatures using an almost model independent method described in the following paragraphs. To do this we must assume that PLZJ23, PLZJ93 and PLZJ100 are all single objects, and indeed, we have no evidence of their being binaries.

With the $I$, $Z$, $J$, $H$, $K$, [3.6] and [4.5] filters measured magnitudes, it is possible to sum the corresponding fluxes, add a small modelled contribution from the wavelengths not covered, and thus, using the distance to the Pleiades, obtain the bolometric luminosity. If we further assume a suitable brown dwarf radius, we can then determine the effective temperature.
We have used the distance to the Pleiades (133\,pc; \citealt{an07}), the AMES-COND spectra \citep{allard01}, M$_{K}$ and the age of the Pleiades (125\,Myr; \citealt*{stauffer98}) to calculate the brown dwarf radii, R.

We then converted our Vega magnitudes to intensities, using the standard zero magnitude intensities \citep*{bessell98}. These were then multiplied by the known filter widths to obtain fluxes. These fluxes were then summed and multiplied by 4$\pi$d$^{2}$ where d is 133 pc (the distance to the Pleiades) to get most of the bolometric luminosity.  

The above paragraph describes the principle of the method and also provides a first T$_{\rm eff}$ estimate, using the method below. However, the Vega spectrum is very different from that of a brown dwarf. Thus for each filter we integrate the Vega spectrum over the filter transmission profile. We also integrate an AMES-COND spectrum with the first estimate T$_{\rm eff}$ over the filter profile. Comparing the two gives the normalisation factor of the AMES-COND spectrum. Finally, we integrate the AMES-COND spectrum over the wavelength region where the transmission exceeds 5 per cent to calculate the total filter flux. As above, then summing the filter fluxes and multiplying by 4$\pi$d$^{2}$ gives the majority of the bolometric luminosity.

To get the total luminosity we then have to add a very small contribution for the fluxes that lie between the $I$, $Z$, $J$, $H$, $K$ and the [3.6] and [4.5] filters and also a contribution for the flux redwards of the [4.5]\,\mic filter (i.e. $>$5.0\,\mic). To do this, we make an estimate of the effective temperature and used the AMES-COND model spectra \citep{allard01} to obtain the surface brightness in the above regions. We then sum these surface brightnesses and multiply by the surface area of the brown dwarf using the radii given in the COND model isochrones \citep{baraffe03}. This contribution is then added to the observed luminosity to obtain the total luminosity. We find, depending slightly on which dwarf is measured, that $\approx$90 per cent of the total luminosity is observed and only $\approx$10 per cent is derived from the models.

We then used the relationship between R/R$_{\odot}$, L/L$_{\odot}$ and T$_{\rm eff}$/T$_{\rm eff \odot}$ to calculate the effective temperature. We then iterated the calculation using the calculated effective temperature instead of the estimated one when calculating the surface brightness from the model spectra.

Our results are an effective temperature of  1121$\pm$50 K and 1252$\pm$50 K, for PLZJ93 and PLZJ23 respectively (we shall assume values of $\approx$1100 and $\approx$1200 K hereafter). The main source of error on this calculation is from the measured photometry (which is in general 10 per cent photometry). A further source of error due to the assumed model radius is given in the following section.
 As we have no $Spitzer$ magnitudes or $H$ band for PLZJ100, we cannot
estimate its T$_{\rm eff}$ other than to say its broadband and methane colours mean is is likely cooler than PLZJ93, and so we can give an upper limit on the effective temperature of $\approx$1100\,K. 
These temperatures correspond to masses between 11 and 12 M$_{\rm Jup}$ using the COND models for 120 Myr \citep{baraffe03}.

\section{Discussion}

The estimated temperatures and the COND models for 120\,Myr can be used to estimate the amount of $H$ band methane absorption we expect in these objects. Figure \ref{models} shows both objects are expected to have  
-1.0 $>CH_{4s}-CH_{4l}>$ -0.5. Both PLZJ93 and PLZJ100 are observed to have positive methane indices however, indicating no methane absorption. Thus low gravity
seems to have a greater effect on the L-T transition than the models predict.

How robust is this result? Clearly it depends on our measured effective temperatures being correct. If T$_{\rm eff}$ was in fact as high as 1400 K, then we
would not expect a negative methane colour index. As explained above, the L/L$_{\odot}$ values only depend on the models at a 10 per cent level. \citet*{saumon08} have 
pointed out that the structural parameters L/L$_{\odot}$ and R/R$_{\odot}$ show good agreement between models by various authors. Also, the modelled R/R$_{\odot}$
 varies very slowly with mass, only 2 per cent change is seen in the radius for a factor of 4 change in mass at the age of the Pleiades.
If we add a $\pm$20 Myr uncertainty (which is larger than the limits given by \citealt{stauffer98}) on the age of the Pleiades, then this introduces a $\pm$ 5 per cent
 error in  R/R$_{\odot}$ and thus a 2 per cent error on the effective temperature. We therefore  believe our estimates of T$_{\rm eff}$ are robust.
The COND model isochrones \citep{baraffe03} give a log g = 4.3 for both PLZJ93 and PLZJ100. Old (5\,Gyr) field dwarfs of the same effective temperature (T$_{\rm eff} \approx$1100\,K) have a log g = 5.4. This leads us to be confident that lowering the gravity by a factor of 10 has a dramatic effect on methane absorption, reducing the field dwarf methane colour index from $\approx$-0.8 to a small, but positive value.

We now compare PLZJ93 with field dwarfs of similar effective temperatures at other wavelengths. \citet{warren07} find a relationship between effective temperature and  $H$-[4.5] colour for field dwarfs. Using this relationship we have determined that at 1100\,K, the effective temperature of PLZJ93, $H$-[4.5] = 2.1. Taking data from \citet{patten06}, Table \ref{pat} gives the colours of several field dwarfs (2MASSIJ0755480+221218, 2MASSIJ2339101+135230, 2MASSIJ2356547-155310, 2MASSIJ1534498-295227, 2MASSIJ1546291-332511) at this $H$-[4.5] colour, which corresponds to spectra types T5-T5.5,  an average T dwarf calculated from these field dwarfs, and the colours of PLZJ93. The bottom two lines of Table \ref{pat} allows us to compare PLZJ93 with an average field dwarf of the same effective temperature.

\begin{table*}
\caption{\label{pat}Colours for field dwarfs with $H$-[4.5] $\approx$ 2.1, and an average T dwarf composed of the 5 2MASS objects. PLZJ93 is also included in the table. The 2MASS photometry has been converted onto the MKO filter system using the spectral type dependant conversions presented in \citet*{stephens04},  as an empirical sequence of field dwarfs.}
\begin{center}
\begin{tabular}{l c c c c c c}
\hline
Name&Spectral type&$H$-[4.5]&$J-H$&$J-K$&$J-[3.6]$&$J-[4.5]$\\
\hline

2MA0755+2212& T5&2.294&-0.553&-0.769&0.651& 1.741\\
2MA2339+1352& T5.0 &2.094 &-0.503& -0.729& 0.721 &1.591\\
2MA2356-1553& T5.0& 2.054& -0.533& -0.619&0.521& 1.521\\
2MA1534-2952& T5.5& 2.072&-0.475&-0.721& 0.676 &1.596\\
2MA1546-3325& T5.5&2.102&-0.155&-0.271&1.116&1.947\\
Average& T5&2.123&-0.444& -0.622& 0.737& 1.679\\
PLZJ93&-&3.026&0.013&0.548&2.702&3.039\\
\hline
\end{tabular}
\end{center}
\end{table*}

As expected, $J$-$H$ is redder for PLZJ93 since it shows no methane absorption, and so is brighter in the $H$ band. $J$-$K$ is also redder for PLZJ93, meaning that it is brighter in the $K$ band than the field dwarf. Methane absorption and collision induced molecular hydrogen absorption (H$_{2}$CIA) both occur in the $K$ band and the H$_{2}$CIA in particular is very sensitive to pressure and gravity \citep{borysow97}. This H$_{2}$CIA is weak for low gravity dwarfs, which when combined with the H$_{2}$O opacity, which also decreases with gravity combines to change the profiles of the $H$ and $K$ band fluxes \citep{mohanty07}, making the $H$ band more peaked and the $K$ band brighter.  This effect in the $H$ band was first seen by \citet*{lucas00} in cool brown dwarf candidates in the Trapezium, and has also been noted in spectra of Pleiades L dwarfs  by \citet{bihain10}.
We thus suggest that the brighter $K$ band magnitude for PLZJ93 is due to the combination of reduced opacity from methane and H$_{2}$CIA in a low gravity atmosphere which contributes to the red $H-K$ and $J-K$ colours seen here.

\begin{figure}
\begin{center}
\scalebox{0.35}{\includegraphics[angle=270]{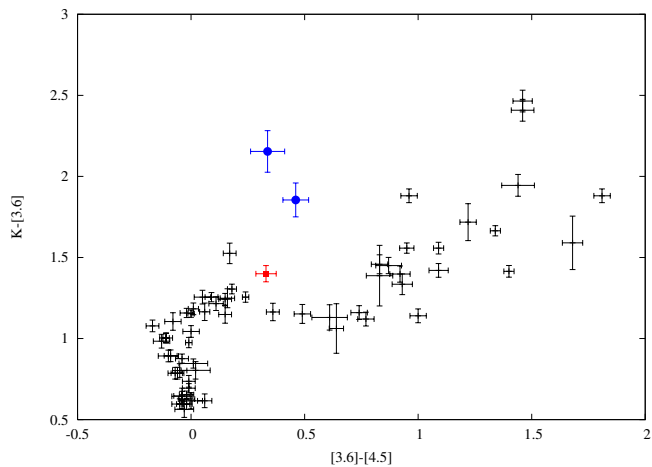}}
\caption{\label{k36}[3.6]-[4.5], $K$-[3.6] colour-colour diagram. PLZJ93 and PLZJ23 (blue circles) and the low gravity  dwarf HN Peg B(red square) are plotted with the field dwarfs from \citet{patten06} (black) providing an empirical L and T dwarf sequence. The $K_{s}$ magnitudes for the Patten objects were converted to the $K$(MKO) using the spectral type dependent conversion in \citet*{stephens04} It can be seen that the lower gravity objects sit above the empirical sequence for field dwarfs.}
\end{center}
\end{figure}

There is another methane band present in the [3.6] IRAC filter and so we expect PLZJ93 to be bright in this filter as well. This proves, quite dramatically, to be the case with a $J$-[3.6] colour difference of $\approx$2 between PLZJ93 and the field comparison. Figure \ref{k36} shows the $K$-[3.6], [3.6]-[4.5] two colour diagram for field stars from \citet{patten06} and PLZJ93 and PLZJ100. The two Pleiads stand very clearly above the field star sequence.

Also shown in Figure \ref{k36} is the low gravity T dwarf HN Peg B.  HN Peg B, has a spectral type of T2-3 and if it were a field dwarf, it would have an effective temperature of 1200-1400 K and a log$\approx$5.0. However, using the age of the system, which is relatively well constrained at 300$\pm$200\,Myr, and evolutionary models, the effective temperature is calculated to be $\approx$1000\,K \citep{leggett08}. Interestingly, HN Peg B does show the expected amount of methane absorption. Using the IRTF Spex infrared spectrum presented in \citet{luhman07}  and integrating over the methane filter profiles, we find  $CH_{4s}-CH_{4l}$ = -0.42. This is consistent with a T2-3 dwarf (as one would expect as the spectrum was used to determine the spectral type). A comparison of the Li measurements for HN Peg A with local open clusters indicates that the system is older than the Pleiades \citep{chen01}, despite the error bar on the age encompassing 100  Myr. This suggests that while the mid-IR colours at [3.6] and [4.5] microns are affected by low gravity up to the age of 300 Myr, the effects of low gravity on methane absorption occur at much younger ages.
\citet{burgess09}, comment that their T dwarf candidate discovered in IC348 has a roughly balanced [3.6]-[4.5] colour as lower gravity means an increase of CO absorption (and hence a decrease of CH$_{4}$ absorption) in the [4.5] micron band. Our Pleiads (and HN Peg B) do not show this ([3.6]-[4.5] $\approx$0.3-0.4; IC348\_CH4\_2 [3.6]-[4.5] $\approx$0.9) or the blue K-[3.6] colour as seen in the IC348 object, indicating that they are of an earlier spectral type with less methane absorption. 
It is clear, however that the $K$-[3.6], [3.6]-[4.5] two colour diagram is an excellent diagnostic for low gravity and hence youth.

The $J$-[4.5] colour also shows an excess for PLZJ93 over the template field dwarf. The reason for this is not obvious. The [4.5] filter covers the CO band, with the band
edge at 4.6 microns. No methane in the object should mean more CO (for a fixed amount of C), and therefore more absorption. Recently \citet*{yamamura10} have seen evidence of CO and CO$_{\rm 2}$ in the [4.5] filter in L dwarfs and it is stronger than predicted by models. In the past the CO has been attributed to vertical mixing in the atmospheres \citep{oppenheimer98}, however, it remains to be seen  if low gravity is affecting the CO$_{\rm 2}$ and thus removing this source of absorption.


The $H$-[4.5] colour of PLZJ93 (T$_{\rm eff}$=1107\,K) is 3.03 which does not fit the \citet{warren07} relation for field stars cooler than 1000 K. This relation was also confirmed by \citep{leggett10} who also noted that it became stronger below 800 K, but used the low scatter in the relationship to determine that it is not gravity sensitive in this temperature regime. At higher temperatures, such as for PLZJ93, the relation is quite likely to be gravity/age sensitive.

PLZJ23, PLZJ93 and PLZJ100 are all clearly cooler than field dwarfs with spectral type T0, but none of them have any methane absorption as judged by their methane colour index.
We assume PLZJ23 has a positive methane colour index, but due to its broad band spectral type being estimated at L8, it was not observed in methane filters. The definition of the 
T spectral class is the appearance of methane so strictly, via this definition, PLZJ93 and PLZJ100 cannot be described as T dwarfs. The colours of these two objects are, however, not compatible with those of field L dwarfs as presented in \citet{leggett02}. 
 Our colours for these Pleiades dwarfs also do not appear to be consistent with the colours types presented in \citet{cruz09} for a low gravity L dwarf sequence up until L5. The colours of a low gravity L5 dwarf are given as $J-H=$1.52$\pm$0.04 and $J-K_s=2.52\pm0.03$. Our objects are much bluer that this. It may however be the case that low gravity late L dwarfs show notably different colours to high gravity late L dwarfs which may explain this discrepancy. This may be an additional indication that the L-T transition occurs at a lower temperature in low gravity objects than for high gravity objects.
It does appear however,  that until an L dwarf sequence is available for low gravity late L and early T dwarfs we have a classification problem and cannot assign these Pleiads an estimated spectral type based upon their photometry.

If spectra could be observed for PLZJ23, PLZJ93 and PLZJ100 it might be possible to classify them as late low gravity L dwarfs, i.e. an extension of the low gravity sequence started in \citet{bihain10} for the Pleiades, and \citet{cruz09} for field dwarfs.

\section{Summary}
Using methane and \textit{Spitzer} IRAC imaging we have shown that 3 of the 6 L-T dwarf candidates discovered by \citet{casewell07} are
likely galaxies. Of the remaining 3, all have photometry that shows that they are likely members of the Pleiades open star cluster and are
low gravity L and T dwarfs. Our findings are consistent with those of \citet{burgess09} on IC348 and \citet{bihain10} on L dwarfs in the Pleiades
as we also see a redder $J-K$ colour and an inhibited methane absorption in these low gravity Pleiads compared to field dwarfs.
We have also discovered a problem with the classification of these objects as we have shown their atmospheres contain no methane, meaning they cannot be described as T dwarfs. They are, however, not L dwarfs, as their broadband colours show. We thus have a classification problem and are unable to assign these objects a (photometrically based) spectral type.

\section{Acknowledgements}
The authors would like to thank the anonymous referee for their useful comments which have improved this manuscript. 
We would also like thank France Allard for the use of the Phoenix web simulator at http://phoenix.ens-lyon.fr/simulator/index.faces which has been used in this research.Many thanks also go to  Kevin Luhman for making his spectrum of HNPegB available to us.
This work is based on observations obtained at the Gemini Observatory (GN-2007B-Q-42), which is operated by the
Association of Universities for Research in Astronomy, Inc., under a cooperative agreement
with the NSF on behalf of the Gemini partnership: the National Science Foundation (United
States), the Science and Technology Facilities Council (United Kingdom), the
National Research Council (Canada), CONICYT (Chile), the Australian Research Council
(Australia), Ministério da Ciência e Tecnologia (Brazil) 
and Ministerio de Ciencia, Tecnología e Innovación Productiva  (Argentina).
The United Kingdom Infrared Telescope is operated by the Joint Astronomy Centre on behalf of the Science and Technology Facilities Council of the U.K.
This work is also  based [in part] on observations made with the Spitzer Space Telescope, which is operated by the Jet Propulsion Laboratory, California Institute of Technology under a contract with NASA.
This research has made use of NASA's Astrophysics Data System Bibliographic Services.

\bibliographystyle{mn2e}

\bibliography{mnemonic,plei}
\label{lastpage}

\end{document}